\documentclass[aps,twocolumn,showpacs,superscriptaddress,groupedaddress, nofootinbib]{revtex4-1}

\usepackage{hyperref}
\usepackage{multirow}
\usepackage{mathrsfs,amsmath} 
\usepackage{verbatim}
\usepackage{amsmath}
\usepackage{amsfonts}
\usepackage{amssymb}
\usepackage{amsthm}
\usepackage{bm}
\usepackage{xparse}
\usepackage{xcolor}
\usepackage{textcase}
\usepackage{url}
\usepackage{lipsum}
\usepackage{appendix}
\usepackage{dsfont}
\usepackage{epstopdf}
\usepackage{footnote}
\usepackage{tgbonum}
\usepackage{blindtext}
\usepackage{enumitem}
\usepackage{mathrsfs,amsmath} 
\usepackage{float}
\usepackage{subfloat}
\usepackage{wrapfig}
\usepackage{pslatex}
\usepackage{amsmath}
\usepackage{amsfonts}
\usepackage{setspace}
\usepackage{epstopdf}
\usepackage{wrapfig}
\usepackage{csquotes}
\usepackage{hyperref}
\usepackage{graphicx}
\usepackage{amssymb}
\usepackage{multirow}
\usepackage{xcolor}
\usepackage{framed}
\usepackage{mathtools}
\usepackage{tcolorbox}
\definecolor{shadecolor}{rgb}{0.8,0.9,1}
\DeclareDocumentCommand{\Tr}{m m O{\big}}{{\rm Tr}_{\:\!{#1}}#3({#2}#3)}

\newcommand{\R}{\mathbb{R}}
\newcommand{\Q}{\mathbb{Q}}
\newcommand{\N}{\mathbb{N}}

\begin{document}
\title{Indeterminism, causality and information: Has physics ever been deterministic?}
\author{Flavio Del Santo}
\affiliation{
Institute for Quantum Optics and Quantum Information (IQOQI-Vienna), A-1090 Vienna, Austria; and
Faculty of Physics, University of Vienna, A-1090 Vienna, Austria; and
Basic Research Community for Physics (BRCP)}
%\author{Nicolas Gisin}
%\affiliation{Group of Applied Physics, University of Geneva, 1211 Geneva 4, Switzerland
%}

\date{\today}

\begin{abstract}
A tradition handed down among physicists maintains that classical physics is a perfectly deterministic theory capable of predicting the future with absolute certainty, independently of any interpretations. It also tells that it was quantum mechanics that introduced fundamental indeterminacy into physics. We show that there exist alternative stories to be told in which classical mechanics, too, can be interpreted as a fundamentally indeterministic theory. On the one hand, this leaves room for the many possibilities of an open future, yet, on the other, it brings into classical physics some of the conceptual issues typical of quantum mechanics, such as the measurement problem. We discuss here some of the issues of an alternative, indeterministic classical physics and their relation to the theory of information and the notion of causality.
\end{abstract}

\maketitle
%%%%%%%%%%%%%%%%%%%%%%%%%%
\section{When did physics become unpredictable?}

Like all of the  human activities, also science maintains traditions that are handed down from generation to generation and help to form the identity of a community.
%These traditions often take the form of romantic stories and myths, featuring heroes and their epic battles.
One specific story that seems to have crystallized among practitioners is that classical physics (i.e., Newton's mechanics and Maxwell's electrodynamics) would allow, in principle, to predict everything with certainty. The standard story continues by telling that the foundations of such a theory are  perfectly well understood and  free of any interpretational issues. In particular, it is widely accepted that classical physics categorically entails a \emph{deterministic} worldview.

Indeed, due to the tremendous predictive success of Newtonian physics (in particular in celestial mechanics), it became customary to conceive an in principle limitless predictability of the physical phenomena that would faithfully reflect the fact that our Universe is governed by determinism. This view was advocated and vastly popularized in the early nineteenth century by Pierre-Simon Laplace, who envisaged the possibility for a hypothetical superior intelligence --which went down in history as \emph{Laplace's demon}-- to predict the future states of the universe with infinite precision, given a sufficient knowledge of the laws of nature and the initial conditions \cite{laplace}:
\begin{displayquote}
Given for one instant an intelligence which could comprehend all the forces by which nature is animated and the respective situation of the beings which compose it --an intelligence sufficiently vast to submit these data to analysis-- it would embrace in the same formula the movements of the greatest bodies in the universe and those of the lightest atom; to it nothing would be uncertain, and the future as the past would be present to its eyes.
\end{displayquote}

The standard story goes on by stating that this faith in perfect determinism was abruptly shattered by the advent of quantum theory, which, with its probabilistic predictions, made indeterministic doubts burst into physics for the first time.\footnote{By indeterminism we denote the sufficient condition that there exists at least one phenomenon, or a type of phenomena, which does not obey determinism.} But is it really so? 
In this essay, we will show that this is not necessarily the case and that the alleged fundamental difference between classical and quantum physics based on their alleged inherently deterministic, respectively indeterministic, character should be rethought. 

From the historical point of view, as early as 1895 --thus before that any quantum effect was discovered and such theory formulated-- the father of the kinetic theory of gases, Ludwig Boltzmann, already doubted the very possibility of having perfect determinism at the microscopic scale (see, \cite{jammer73, delsanto}). It ought to be noticed that according to the standard understanding of classical statistical mechanics, probabilities are there introduced to account for a lack of knowledge about the actual state of affairs (epistemic randomness) and are not supposed to be \emph{irreducible}.\footnote{Probabilities are said to be irreducible if ``it is not possible by further investigation to discover further facts that will provide a better estimate of the probability.'' \cite{dowe}.} This is due to the fact that statistical physics deals with an enormous amount of components (and of degrees of freedom), but it is generally accepted that every single classical particle has a perfectly predetermined behavior (and that in principle this is predictable). Despite this, Boltzmann maintained: ``I will mention the possibility that the fundamental equations for the motion of individual molecules will turn out to be only approximate formulas which give average values.'' \cite{boltzmann}. Also Franz S. Exner, another eminent  Viennese physicists, contemporary of Boltzmann, questioned the validity of determinism in classical physics even at the macroscopic level: ``In the region of the small, in time and space, the physical laws are probably invalid; the stone falls to earth and we know exactly the law by which it moves. Whether this law holds, however, for each arbitrarily small fraction of the motion [...] that is more than doubtful.'' \cite{exner}. One of the intellectual heirs of Boltzmann and Exner in Vienna was the Nobel laureate and founding father of quantum theory, Erwin Schr{\"o}dinger. Indeed, he fully embraced their skeptical positions about determinism: ``As pupil of the venerable Franz Exner I have been on intimate terms for a long time with the idea that probably not microscopic lawfulness but perhaps \lq absolute accident\rq \  forms the foundation of our statistics.'' \cite{schrodinger}.\footnote{This trend of doubting determinism in the Vienna school of statistical physics has been referred to as the \textit{Vienna indeterminism} in the philosophical literature \cite{viennaindet}.}

Interestingly, contrarily to the text-books presentation of classical physics, the fact that classical systems have a perfectly predeterminate dynamics (thus giving rise to perfectly deterministic predictions) is not inherent in the formalism (see Section \ref{formalism}). Rather, it is based on an additional hidden assumption that takes the form of a principle. In a previous work \cite{delsantogisin}, we have named this \emph{principle of infinite precision}. This is articulated in two parts, as follows:
\begin{shaded}
\textbf{\textit{Principle of infinite precision}}\\
\begin{enumerate}
\item \emph{Ontological} -- there exists an actual value of every physical quantity, with its infinite determined digits (in any arbitrary numerical base).
\item \emph{Epistemological} -- despite it might not be possible to know all the digits of a physical quantity (through measurements), it is possible to know an arbitrarily large number of digits.
\end{enumerate}
\end{shaded}
It is only when its formalism is complemented  with this principle that classical physics becomes deterministic. 

However, the principle of infinite precision is  inconsistent with any operational meaning, as already made evident by Max Born. The latter gave pivotal contributions to the foundations of quantum formalism --introducing the fundamental rule that bears his name, which allows to assign probabilities to quantum measurements, and for which he was awarded the Nobel Prizein 1954-- and became critical of determinism, even in classical physics, due to its reliance on ``infinite precision''. Indeed, in his essay \emph{Is Classical Mechanics in fact Deterministic?} \cite{born}, he affirmed: 
\begin{displayquote}
It is usually asserted in this theory [classical physics] that the result is in principle determinate and that the introduction of statistical considerations is necessitated only by our ignorance of the exact initial state of a large number of molecules. I have long thought the first part of this assertion to be extremely suspect. [...] Statements like 'a quantity $x$ has a completely definite value' (expressed by a real number and represented by a point in the mathematical continuum) seem to me to have no physical meaning. [Because they] cannot in principle be observed.
\end{displayquote}

To explain how infinite precision and determinism relate to one another it is interesting to rephrase a simple example devised by Born.
%%%%%%%%%%%%%%%%%%%%%%%%%%
\begin{figure}[h!]
\includegraphics[width=8.5cm]{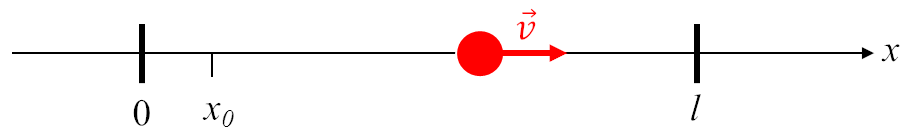}
\caption{Toy example of a system (a classical particle confined in a cavity), in which the indeterminacy on the initial conditions is amplified in the future indeterminacy of the physical state while time passes.}
\label{born}
\end{figure}
%%%%%%%%%%%%%%%%%%%%%%%%%%
Referring to figure \ref{born}, consider a (classical) particle that is bound to move in a one-dimensional cavity with perfectly elastic walls and total length $l$. If the particle has a perfectly determinate (i.e. with infinite precision) initial position $x(t=0)=x_0$ and velocity $\vec{v}(t=0)=\vec{v}_0$, classical physics then allows to predict, also with infinite precision, the future positions $x(t)$ and velocities $\vec{v}(t)$ for any time instant $t$. Yet, imagine to slightly relax the principle of infinite precision and, while $x_0$ remains fully determinate, the initial velocity (say pointing to the right) of the particle has a small indeterminacy, i.e. $v_0 \leq v(t=0)=v_0+\Delta v_0$. Once the particle starts moving, its initial indeterminacy starts to be reflected on the determination of its position at later times. According to the laws of classical mechanics, the range of possible future positions of the particle increases linearly as time passes, i.e. $\Delta x(t)=t \Delta v_0$. This means that, for any arbitrarily small indeterminacy of the initial velocity $\Delta v_0$, there always exists a critical time instant $t_c:=l/\Delta v_0$, such that $\Delta x(t=t_c)=l$. Namely, independently of how small is the initial indeterminacy, it is sufficient to wait enough long time for having complete indeterminacy on the particle's position within the cavity. This clearly shows that the principle of infinite precision is a necessary condition for determinism.

This example is one of the simplest instantiations of those systems whose future dynamics is highly susceptible to a variation of the initial conditions (a property called \emph{instability}). Such a phenomenon is typical of the so-called \emph{chaotic systems} wherein the uncertainty in the determination of future values of some physical quantities increases exponentially with elapsed time. Actually, our example displays  what in chaos theory is referred to as the \emph{real butterfly effect} \cite{butterfly}. Namely, not only high sensitivity to the initial conditions, but --since the system is bounded, and, accordingly, it is the region of phase space within which its physical state can evolve-- after a certain time the uncertainty saturates the whole allowed region of phase space. This means that after a certain critical time the distribution of the states in phase space is the homogeneous one.\footnote{We acknowledge Sabine Hossenfelder's essay ``Math Matters'' in the 2020 FQXi Essay Contest for Ref. \cite{butterfly}.} As a matter of fact, criticisms of classical determinism became more severe in the second half of the last century, when the chaos theory was further developed and its fundamental consequences understood \cite{ornstein, prigogine}. 

Furthermore, the challenges to determinism in classical physics experienced a revival in very recent years, when several scholars formalized the fact that predetermined physical quantities seems to be at odds with information-theoretic arguments \cite{dowek, gisin1, NGHiddenReals, blundell, drossel, delsantogisin, lynds}, as we will show in detail in what follows.

%%%%%%%%%%%%%%%%%%%%%%%%%%%%%%%%%%%%%%%%%%%%%%%%%%
\section{The ``orthodox interpretation'' of classical physics}\label{formalism}

%%%%%%%%%%%%%%%%%%%%%%%%%%
\begin{figure*}[]
\includegraphics[width=18cm]{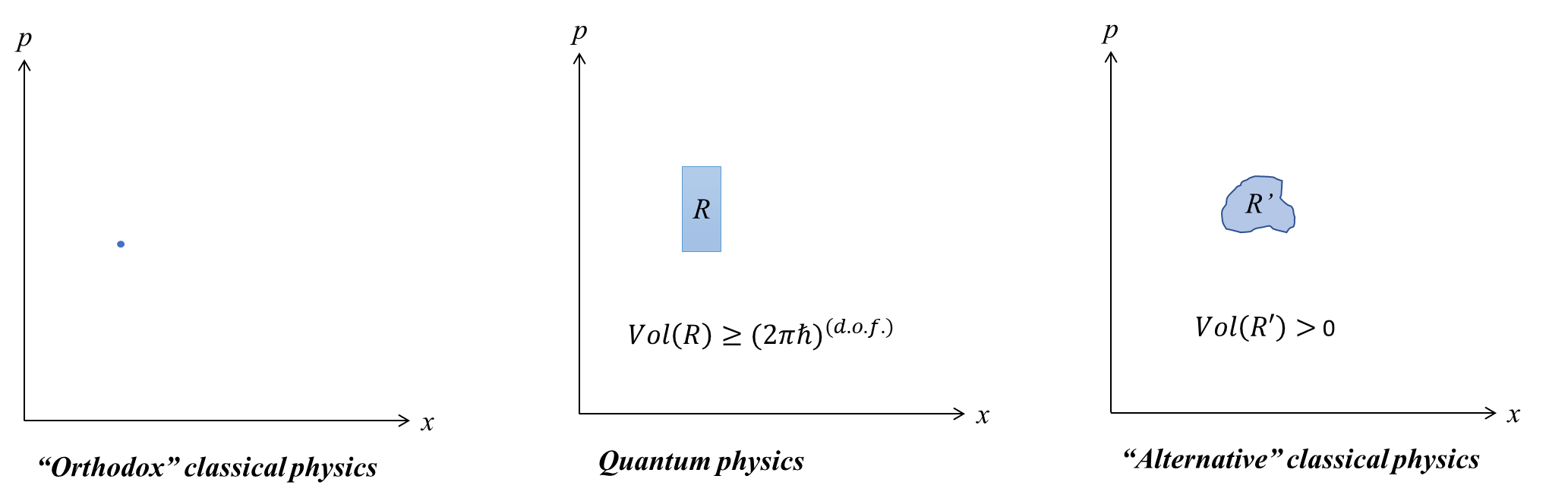}
\caption{\small{Suggestive representations of physical states in phase space, in comparison for ``orthodox'' classical physics (left), quantum physics (middle), and ``alternative'' classical physics (right). In the orthodox classical physics, the state is a mathematical point which determines a unique trajectory (determinism), whereas both in quantum and alternative classical physics the state has a fundamental indeterminacy that leads to an indeterministic dynamics. The peculiarity of quantum physics is that its formalism sets a precise value to the smallest size of a cell in phase space.}}
\label{phasespace}
\end{figure*}
%%%%%%%%%%%%%%%%%%%%%%%%%%
In order to introduce the arguments against the tenability of determinism in classical physics and a possible alternative interpretation thereof,\footnote{We use here the expression ``interpretation'' and not ``theory'' because we consider only empirically indistinguishable predictions (in the same sense of the interpretations of quantum mechanics) \cite{baumann}.} we first ought to recall some pillars of the formalism of that theory. We refer here to the standard formalism together with its metaphysical assumptions (i.e. the principle of infinite precision) as the ``orthodox interpretation'' of classical physics.\footnote{We borrow this name from the foundations of quantum mechanics, where the attribution  ``orthodox'' is usually associated to the most widespread interpretation of the quantum formalism, also called the ``Copenhagen interpretation'', attributed to Niels Bohr and his school.}

Conceptually, classical physics (say Newtonian mechanics, but equivalent arguments apply to classical electromagnetism, too) is characterized by (i) the \emph{physical state} of a system, which accounts for its relevant physical properties and (ii) a set of \emph{general laws} that govern the evolution (backwards and forward in time) of the physical state.\footnote{To these two main aspects of classical physics one has to add a third one, (iii) that there exists a background time which allows to speak about the state of a physical system at a certain instant of time and its evolution at later instants.}
Formally speaking, the dynamical properties of a system are identified by a set of physical quantities, which mathematically are called variables. The collection of these variables (typically position and momentum) is called the \emph{state} of the system and the space composed of all the possible values taken by these quantities is called \emph{phase space}. Moreover, the assumption of the principle of infinite precision results in the fact that classical states are mathematical points in a continuous phase space. Namely, a physical state is an $n$-tuple of real numbers for a $n$-dimensional phase space.

As for the mathematical characterization of  the general laws of classical mechanics, these are ordinary differential equations that take as inputs the values from the state at time $t_0$, called \emph{initial conditions}, and return the state of the system at any arbitrary time $t$. The mathematical theory of differential equation then guarantees that, given any sequence of subsequent states in a certain interval of time, i.e. a \emph{trajectory} in phase space for that interval, there exists always a unique extension thereof, into the past and the future \cite{earman}. This mathematical formalization all together leads, in fact, to the formal definition of \emph{Laplacian determinism}: For any given physical state there exists  a unique evolution, i.e., a unique trajectory in phase space. However, as stressed by Drossel, ``the idea of a deterministic time evolution represented by a trajectory in phase space can only be upheld within the framework of classical mechanics if a point in phase space has infinite precision'' \cite{drossel}.

We have already pointed out how the concept of infinite precision has no operational meaning. This was also recently remarked by Rovelli, who stated  that ``concretely we never determine a point in phase space with infinite precision --this would be meaningless--, we rather say that the system `is in a finite region $R$ of phase space', implying that determining the value of the variables will yield values in $R$.'' \cite{rovelli}. Note that, however, as most of his fellow physicists, Rovelli upholds the ``orthodox'' classical mechanics which considers this a for-all-practical-purposes issue, i.e. at the fundamental level, $R$ degenerates to a mathematical point. In quantum physics, on the other hand, there is a fundamental lower limit to the size of the region in phase space. In particular, the volume of the region, $Vol(R)$, cannot be smaller than the size delimited by the Planck constant (for each degree of freedom), i.e.,
\begin{equation*}
Vol(R)\geq(2 \pi \hbar)^{(d.o.f.)}.
\end{equation*}
And Rovelli refers to this as ``\emph{the} major physical characterization of quantum theory'' \cite{rovelli}.

But are quantum and classical physics necessarily so different on this matters? One has to realize that the principle of infinite precision is not part of the mathematical formalism of classical theory, but rather it belongs to the domain of interpretations. In fact, one can consider --and several arguments point in the direction that one should perhaps do so-- an alternative interpretation of classical physics in which  physical states are not mathematical points characterized by ($n$-tuples of) real numbers. In this way, even classical physics would display a fundamental indeterminacy, and its conceptual difference with quantum mechanics should be scaled down (see Fig. \ref{phasespace}).

Invoking again Born's operationalism, one ought to consider the following \cite{born}:
\begin{displayquote}
A statement like $x = \pi$ cm would have a physical meaning only if one could distinguish between it and $x = \pi_n$ cm for every $n$, where $\pi_n$ is the approximation of $\pi$ by the first $n$ decimals.  This, however, is impossible; and even if we suppose that the accuracy of measurement will be increased in the future, $n$ can always be chosen so large that no experimental distinction is possible. Of course, I do not intend to banish from physics the idea of a real number. It is indispensable for the application of analysis. What I mean is that a physical situation must be described by means of real numbers in such a way that the natural uncertainty in all observations is taken into account.
\end{displayquote}
As we will show in the next section, one can indeed envision an alternative classical physics that maintains the same general laws (equations of motion) of the standard formalism, but dismisses the physical relevance of real numbers, thereby assigning a fundamental indeterminacy to the values of physical quantities, as wished by Born. In fact, ``as soon as one realizes that the mathematical real numbers are “not really real”, i.e. have no physical significance, then one concludes that classical physics is not deterministic.'' \cite{gisin1}.  

%%%%%%%%%%%%%%%%%%%%%%%%%%%%%%%%%%%%%%%%%
\section{An alternative, indeterministic interpretation of classical physics}

%%%%%%%%%%%%%%%%%%%%%%%%%%%%%%%%%%%%%%%%%
\subsection{Determinism at odds with information principles}

The relaxation of the principle of infinite precision does not come about only as a mere intellectual exercise, or as a proof of principle that classical physics is compatible with alternative interpretations beyond the orthodox one. In fact, the motivation for searching novel interpretations of classical physics stems also from the application of information-theoretic concepts to physics. Indeed, our current understanding tells us that
\begin{displayquote}
information is not a disembodied abstract entity; it is always tied to a physical representation. It is represented by engraving on a stone tablet, a spin, a charge, a hole in a punched card, a mark on paper, or some other equivalent. This ties the handling of information to all the possibilities and restrictions of our real physical world, its laws of physics and its storehouse of available parts. \cite{landauer}.
\end{displayquote}
This view goes under the name of \emph{Landauer's principle}, in short, ``information is physical''.

In Ref. \cite{gisin1}, Gisin gave sound arguments to support the claim that  ``a finite volume of space cannot contain more than a finite amount of information''. Intuitively, this is a direct consequence of Landauer's principle, because each bit of information to be stored requires a certain amount of space, bounded from below by the size of the smallest physical system that can encode it. It is true that today, thanks to the incredible development of the technology of miniaturization, we are able to encode and manipulate information in astonishingly small systems. This allows to reach densities of information storage of about 25 terabytes per centimeter square \cite{tech} on atomic lattices, whereas molecular storage of information in DNA has recently achieved extraordinary densities of information of the order of a million terabyte per cubic millimeter \cite{dna}. 
%In this way, states Emily Leproust --one of the pioneers of DNA-based information storage-- ``you could fit all the knowledge in the whole world inside the trunk of your car''.
These outstanding results notwithstanding, physical systems have a finite size, hence it seems a very reasonable assumption to believe that there is a finite limit to the possible information density. 

Furthermore, a well known formal theoretical argument sets a limit to the allowed information density, called \emph{Bekenstein bound} \cite{bound}, states that the information $I$ (in number of bits) contained in a system circumscribed by a sphere of radius $R$ is smaller than the mass-energy $E$ enclosed in the same sphere, i.e., 
\begin{equation*}
\frac{I}{2\pi R} \leq \frac{E} {\ln2},
\end{equation*}
where we have adopted the Planck units (i.e., $c = \hbar =1$). The intuition behind this is that the storage of each bit of information is associated with a certain amount of energy and that unbound densities of energy degenerate into black holes.

Coming back to the orthodox interpretation of classical physics, we have already shown how this assumes that physical states are mathematical points in phase space, expressed by ($n$-tuples of) real numbers. However, it should be noticed that real numbers contain, in general, an infinite amount of information. As we have learnt since  primary school, the set of real numbers encompasses all the familiar rational numbers and supplement them with the irrational numbers. However, even among the irrational numbers, there are fundamental conceptual differences that have relevant consequences for the role they are attributed in physics. All the irrational numbers we are used to speak about, such as $\sqrt2$ or $\pi$, are, in fact, \emph{computable} irrational numbers. This means that they can be compressed into an algorithm of finite length which, at every iteration, outputs the next digit of the considered number. For instance, an algorithm (but not the only one) to construct $\pi$ is given by  computing (each digit of) the ratio of the circumference of any circle to its diameter. So, although an irrational computable number has infinite digits without a periodic pattern, and, as such, it would take infinite time (i.e., iterations of the associated algorithm) to get all the digits, its actual information content is finite. Everything there is to know about it is contained in the algorithm that generates it. More precisely, the (finite) information content of a computable number corresponds to the amount of information in bits of shortest algorithm that outputs that number (i.e., its \emph{Kolmogorov complexity}).  What is however disconcerting, is that the amount of computable numbers among all the real numbers is infinitely small (i.e., it forms a subset of Lebesgue measure zero). Technically, the probability of picking a computable number from the set of real numbers is zero (see also \cite{gisin1, dowek}).

Putting together the above arguments, we come to the conclusion that real numbers cannot be physically meaningful insofar as their information content is almost always infinite. One thus ought to consider alternative interpretations of classical physics that do not enforce the principle of infinite precision. Namely, interpretations that do not assume that physical quantities take values in the real numbers. Note again that without real numbers, one cannot any longer uphold determinism in classical physics.

In this view, the orthodox interpretation of classical physics can be regarded as a deterministic completion of an indeterministic model, in terms of \emph{hidden variables}: Namely, the real numbers \cite{gisin1, NGHiddenReals}. This is reminiscent of Bohm's \cite{bohm} or Gudder's \cite{gudder} hidden variable models of quantum physics, which provide a deterministic description of quantum mechanics by adding (in principle inaccessible) supplementary variables, whereas the orthodox interpretation takes probabilities (therefore indeterminism) to be irreducible.

%%%%%%%%%%%%%%%%%%%%%%%%%%%%%%%%%%%%%%%%%
\subsection{``Finite information quantities'' (FIQs).} 
\label{fiqs}
To overcome the problem of the infinite information content of real numbers in the context of physics, an explicit alternative model has been sketched in Ref.  \cite{gisin1} and developed in greater detail in Ref. \cite{delsantogisin}. This model entails an alternative indeterministic interpretation of classical mechanics. We review here its main features.

In the spirit of the previous considerations, let us leave the dynamical equations of Newtonian mechanics unchanged, but let us relax the principle of infinite precision by substituting the real numbers with newly defined quantities. We refer to them as ``finite-information quantities'' (FIQs), which, while providing the same empirical predictions as the orthodox interpretation of classical physics, have no overlap with real numbers (they are not a mathematical number field, nor a proper subset thereof). 

Let us start by considering again the orthodox interpretation. Let a physical quantity $\gamma\in \R$ (say the position of a particle moving in one dimension) lie, without loss of generality,  in the interval $[0,1]$ and write it in binary base:
\begin{equation*}
\gamma=0.\gamma_1\gamma_2\cdots \gamma_j \cdots,
\end{equation*}
where $\gamma_j\in\{0,1\}$, $\forall j\in \N^+$. This means that, being $\gamma \in \R$, its infinite bits are \emph{all} given at once, i.e., always determined.

Consider now the following alternative model to describe physical quantities which introduces an element of randomness in such a way to always guarantee the finiteness of the information content.  We thus introduce the following:
\begin{shaded}
\textbf{Definition - \textit{propensities}}\\
There exist objective properties, named \emph{propensities}, $q_j\in [0,1] \cap \Q$, for each digit $j$ of a physical quantity. A propensity quantifies the tendency of the $j$th binary digit to take the value 1.
\end{shaded}
The concept of propensities, borrowed from Popper's objective  interpretation of probabilities \cite{popper}, can be understood from the limit cases, namely when they are either 0 or 1. For example, $q_j=1$ means that the $j$th digit will take value 1 with certainty. On the opposite end, if a bit has an associated propensity of 1/2, it means that the bit is totally indeterminate. We posited that propensities are rational numbers, but in general it is enough that their information content is always finite (e.g., they could be computable real numbers). In order to define physical quantities, we thus have to define the following:
\begin{shaded}
\textbf{Definition - \textit{FIQs}}\\
A \emph{finite-information quantity} (FIQ) is an ordered list of propensities $\{ q_1, q_2, \cdots , q_j, \cdots \}$, each associated to a bit of a physical quantity, such that the overall information content is finite, i.e.,  $\sum_j I_j < \infty$, where $I_j$ is the information content of the$j$-th propensity (as expressed by some reasonable measure).
\end{shaded}
Note that a previous work \cite{delsantogisin}, we have suggested a straightforward way to construct a FIQ, i.e. to assume that after a certain threshold, all the bits to which propensities are associated become completely random, i.e., $\exists M(t) \in \N$ such that $q_j = 1/2, \forall j>M(t)$. In this way, the propensities are all independent and it is possible to choose $I_j=1-H(q_j)$, where $H$ is the binary entropy function of its argument. It's trivial to check that in that scenario  $\sum_j I_j < \infty$. However, in Ref. \cite{callegaro}, it was pointed out a weakness of the simple latter scenario, namely that the mutual independence between propensities of a FIQ is not preserved under under a basic operation such as a change of unit. We have shown in  \cite{reply} that this does not jeopardizes the FIQ program, but indeed forces us to introduce more complex ways to construct FIQs, such that correlations between propensities are properly introduced. Despite the criticism in \cite{callegaro}, for conceptual simplicity we still present in what follows the simple model to construct FIQs using independent propensities, because this will more intuitively allow to discuss the main conceptual novelties and issues. 

In general, since we require our alternative interpretation to be empirically equivalent to the orthodox one, at least the digits of a physical variable that are already known (i.e., measured) at time $t$ should be fully determined. Therefore, the propensities of the first, more significant, $N(t)$ digits  should be already actualize, i.e. $q_i \in\{0,1\}, \forall i\leq N(t)$. 
We are now ready to express a physical quantities $\gamma$ in this FIQ-based interpretation:
\begin{equation*}
\gamma \left(N(t), M(t)\right)=0.\underbrace{\gamma_1\gamma_2\cdots \gamma_{N(t)}}_{\textrm{determined }\gamma_j\in \{0, 1\}} \overbrace{?_{N(t)+1}\cdots ?_{M(t)}}^{?_k\textrm{, with } q_k\in(0, 1)}\underbrace{?_{M(t)+1}\cdots}_{?_l\textrm{, with } q_l=\frac{1}{2}},
\end{equation*}
where the symbols $?_i$ means that the digit in position $i$ is not yet determined.
Notice that in this framework  the potential property of becoming actual (a list of propensities, FIQ), has somehow a more fundamental status than an already actualized value (a list of determined bits). In fact, in this alternative interpretation, a state would be the collection of all the FIQs associated with the dynamical variables (i.e., the list of the propensities of each digit). Thus, even two systems that are to be considered identical at a certain instant of time (in the sense that they are in the same state) will have, in general, different actual values at later times.  But then, how does the actualization happen in such a way that is compatible with the observed results? To answer this question we need to discuss what is a measurement in a non-deterministic physics.

%%%%%%%%%%%%%%%%%%%%%%%%%%%%%%%%%%%%%%%%%
\subsection{The classical ``measurement problem''}

Any indeterministic interpretation of a physical theory needs to face the questions (\emph{measurement problem}): How does a single value of a physical variable become actualized out of its possible values? Or how does potentiality become actuality? Our experience, in fact, tell us that every time a quantity gets measured there is only one value registered by the instrument. In order to address this issue, however, it is necessary to first ask: What is a measurement? This long-lasting question is one of the most profound open problems of the foundations of quantum physics (see, e.g. \cite{brukner}). As we have recalled, the latter is normally considered the first theory to have introduced fundamental indeterminacy in the domain of physics. Yet, as soon as an indeterministic interpretation of classical physics is upheld, this is also subject to a measurement problem.

Let us operationally define what are the minimal requirement for a process to be considered a measurement:
\begin{shaded}
\textbf{Definition - \textit{Minimal requirements for a measurement}}\\
\begin{enumerate}
\item \emph{Stability}: Consecutive measurements of the same quantity leave the already determined digits unchanged.
\item \emph{Intersubjectivity}: Different agents can access the same measurement outcomes.
\item \emph{Precision improvability}: With more accurate measurement apparatuses, more digits become available (with the former two properties).
\end{enumerate}
\end{shaded}

As for stability, it should be remarked that it is of course possible that the dynamical evolution would change the state of the system under consideration, and therefore the outcomes of measurements occurring at two (arbitrarily distant) instants of time. However, what is assumed here is a trivial evolution, or equivalently a short enough time interval between consecutive measurements, thus focusing only on the changes of the states due to measurements. This assumption is customarily upheld in explaining quantum mechanics, when one says that two consecutive measurements of the same observable yield the same results (obviously if performed in the same basis).

In order to be an empirically adequate model, in the FIQ-based indeterministic interpretation of classical physics, too, one needs to explain how to comply with these properties of measurements. By construction of FIQs, propensities are objective properties subjected to fundamental irreversibility (i.e., once they become either 0 or 1 the remain unchanged), and this accounts for the stability and ensures the intersubjective availability of the measurement results.

What is far from being straightforward, however, is the compliance with precision improvability. In Ref. \cite{delsantogisin}, we have introduced two possible ways to account for this property. On the one hand, one can envisage (i) a mechanism that makes the actualization to spontaneously occur as time passes. This resembles the so-called \emph{objective collapse models} of quantum mechanics \cite{grw, gisincollapse, cls}. On the other hand, it is possible to think that (ii) the actualization happens when a higher level requires it, thus, with some \emph{top-down causation} mechanism \cite{topdown}. In this case, it would be the measurement apparatus that ``imposes'' to the physical variables to acquire a determined value. This is clearly reminiscent of the Copenhagen interpretation of quantum mechanics.

Such a ``classical measurement problem'' remains an open problem as much as its more notorious  quantum counterpart. Yet, it should be noticed that however problematic, the fact that both classical and quantum physics share this issue helps to scale down the fundamental difference between these two theories.

%%%%%%%%%%%%%%%%%%%%%%%%%%%%%%%%%%%%%%%%%
\section{(In)determinism and causality}
Like in the orthodox interpretation, in the indeterministic model previously introduced, too, the laws of classical mechanics are taken to be general relations that causally connect physical states at different instants of time. Traditionally, in the philosophy of science, the concepts of determinism and causality have been long wedded, to the extent that usually  Laplacian determinism is often referred to as \emph{causal determinism}. One of the most notable examples of this can be found in Hume, who maintained that a cause is always sufficient for its effect: ``It is not possible on Hume's account, for causes to be less than deterministic.'' \cite{dowe}. Also Leibniz elevated determinism to an \emph{a priori} truth, when formulating his \emph{principle of sufficient reason}: ``There is nothing without a reason, or no effect without a cause'' (quoted in \cite{earman}). And Kant even formulated what is sometimes called the \emph{law of universal causation}, according to which, ``if we [...] experience that something happens, then we always presuppose thereby that something precedes on which it follows in accordance with a rule.'' \cite{kant}.\footnote{Note that Kant refers to the term \emph{rule} as a univocal correspondence and does not contemplate any non-deterministic (e.g., probabilistic) law that relates causes and effects.}

The concept of causation is also traditionally related, at least in science, to the quest for explanation. This means to ask: Why did the (observed) event $E_i$ happen? (here $i$ labels the time at which $E$ occurs). Answering this question in a deterministic worldview seems to us quite meaningless. In fact, were \emph{everything} completely predetermined, this question --like any other one-- would not be a genuine question, in the sense that, for whoever asked it, this was a necessity from the beginning of times. Recalling again Laplace's demon, ``the future as the past would be present to its eyes'' \cite{laplace}. In other words, everything would just be an already shot film that is unrolling, and the script of the film makes you say the line ``why did $E_i$ happen?''. 

Moreover, even if we assume that an agent, or an intellect, is external to \emph{everything} that occurs in the universe (i.e. s/he is really watching the movie from outside), thereby not being included in this predetermined state of the universe, this is not unproblematic. Asking \emph{why} something happened is in this case certainly meaningful but the answer is trivial: Because this is the film I am watching (and there are no other films available!). Again, determinism assumes that given an initial state of the universe and universal laws everything \emph{causally} follows. But this is misleading because there is only one specific initial state and, without alternatives, causation seems a void concept.

On the contrary, indeterminism introduced the possibility of alternatives, thereby making causality meaningful. If one asks the reason why a certain event $E_j$ occurred, is now possible to reply: ``Because another event $E_{i/A}$ happened before (i.e., $i<j$) and not its mutually exclusive alternative $E_{i/B}$''.
Significant progress in weakening the bond between determinism and causality was made in the second half of the nineteenth century, thanks to the work of philosophers the likes of K.R.  Popper \cite{popper}, J. Earman \cite{earman}, W. Salmon \cite{salmon}, P. Dowe \cite{dowe}, H. Reichenbach \cite{reichenbach}, I. J. Good \cite{good} and P. Suppes \cite{suppes2}. Mostly inspired by quantum mechanics, the concept of \emph{probabilistic causality} came about. This maintains that an event $C$ directly influences another event $E$ but is not sufficient for it. A common, and quite grim, example to explain probabilistic causality features the following chain of events (temporally ordered): A scientist, Eric, sits in a sealed room (i.e., without any exchange with the external environment). His colleague, Clara, brings a canister full of radioactive material in Eric's room (ideally, making sure that there are no other exchanges with the environment). While time elapses, the radioactive material will be decaying --at a certain probabilistic rate depending on its chemical composition-- releasing ionizing radiation. Sadly, at some point, Eric develops radiation poisoning. Now,  since decay is governed by quantum mechanics and in that theory probabilities are considered irreducible, there was no deterministic process relating Clara's actions to Eric's condition. However, if you think that Clara can be held accountable for Eric's sickness, then you believe in probabilistic causality.
%%%%%%%%%%%%%%%%%%%%%%%%%%
\begin{figure}[H]
\includegraphics[width=8.5cm]{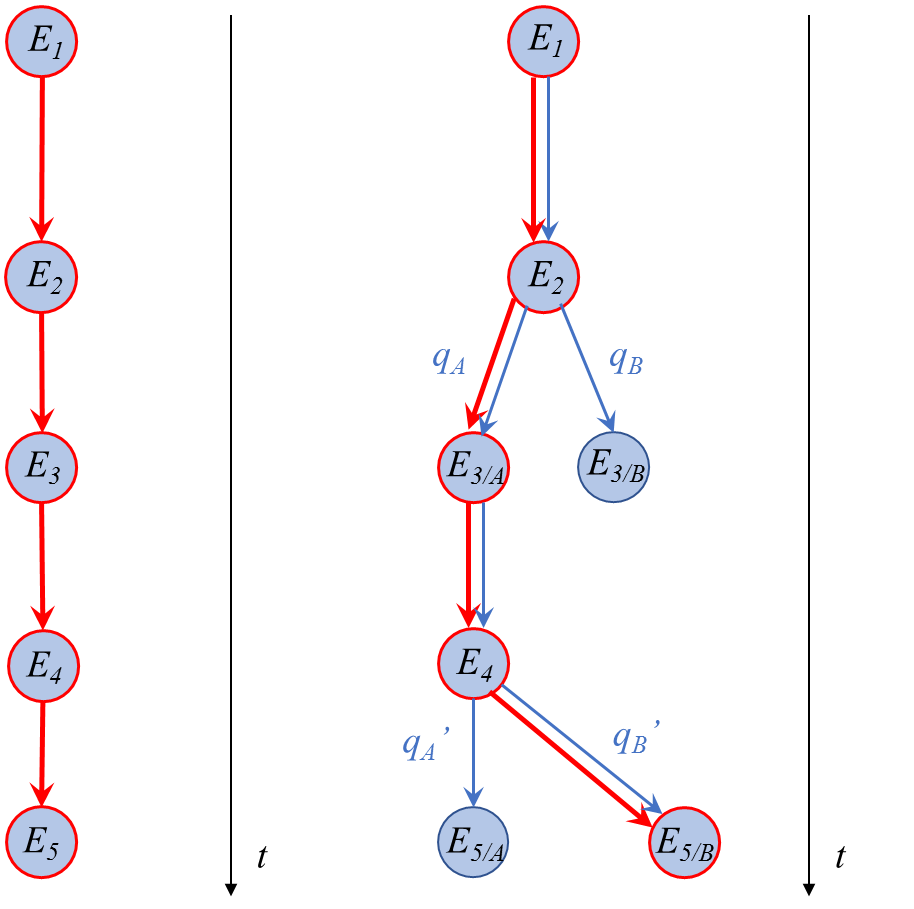}
\caption{Causal graphs of a deterministic (left) and indeterministic (right) series of events. The red arrows relate events that actually happen, while the blue ones denote potential, mutually exclusive, alternative events of which only one will happen with a certain propensity $q_K$ (see main text).}
\label{graph}
\end{figure}
%%%%%%%%%%%%%%%%%%%%%%%%%%
Referring to Fig. \ref{graph}, we can graphically formalize deterministic (on the left) and probabilistic causality (on the right). Both are represented as graphs which are \emph{directed} (causes precede their effects in time), and \emph{acyclic} (an effect cannot be the cause if itself). However, in a deterministic graph, there are no possible alternatives: Everything that can happen does happen.\footnote{Note that it is of course not necessary that each event is effect and cause of one and only one event as in Fig. \ref{graph}-left. We represented this simple chain because we deem less confusing the comparison with the probabilistic graph (Fig. \ref{graph}-right).} On the contrary, a graph representing probabilistic causality (Fig. \ref{graph}-right) is a \emph{multigraph} with two types of edges. The first ones (blue) represent the ``potential causations'' and are \emph{weighted} with the measure or the degree to which an event $E_i$ causes future events $E_{j/K}$, where $i,j$ label the time instants and $K\in \{A, B, ...\}$ the possible mutually exclusive alternatives.\footnote{Note that the celebrated \emph{Many-World Interpretation} of quantum mechanics affirms that all the possible alternative outcomes actually happen, hence refuting the mutual exclusiveness thereof. While this is also a possible further interpretation of the FIQ-based physics, we will not consider this further.} A natural choice for the weights of the potential causation is clearly propensities $q_K$, as defined in Sect. \ref{fiqs}. The second kind of edges (red), instead, represent what actually happened and can be reconstructed in hindsight after the actualization of the potentiality has happened (e.g., after measurements). 

Clearly, the alternative interpretation of classical physics based on FIQs, introduced in section \ref{fiqs}), is causal but not deterministic and can be represented by causal graphs of the second type (Fig. \ref{graph}-right).

Recently, D'Ariano, Manessi and Perinotti \cite{giacomomauro} have pointed out that the notions of determinism and causality are logically independent, namely one can have not only non-deterministic (probabilistic) causal theories, such as quantum mechanics, but in principle non-causal deterministic theories, too. Their argument is carried out in the framework of \emph{operational probabilistic theories}, introduced by some of the same authors in previous works. Without entering the formal details, according to the authors of Ref. \cite{giacomomauro} a theory is said to be causal if there is \emph{no-signaling from the future}. Namely, if the probability of preparing a system in a certain initial state is independent from the choice of measurements that will be performed on the system itself. Determinism  is instead defined as ``the property of a theory of having all probabilities of physical events equal to either zero or one.'' \cite{giacomomauro}. They then cleverly design a toy-theory that is, in fact, deterministic but not causal, according to their definitions.
 
However, defining deterministic behaviors  as a limiting case of probabilistic ones, while is seemingly very natural, leads to subtle issues. Indeed, this boils down to give an interpretation of what probabilities are supposed to mean. If they are taken to have no causal meaning, but being merely measured frequencies of occurrences, then the fact that determinism and causality are logically independent becomes trivial. Consider a typical example of classical correlations: During a vacation in New York, take a pair of shoes and separate them into two identical boxes. Shuffle the boxes in a way that is impossible to know which one contains the left, respectively the right, shoe. Keep one box with you and send the other to a friend in Tokyo. You can now open the box and you figure out that that you kept, say, the left shoe. Then you can infer the following statement with probability one, i.e., deterministically: ``My friend in Tokyo has received a right shoe''. Nobody, however, would ever entail that finding the left shoe in New York has caused the right show to be found in Tokyo.\footnote{One can object that in fact, the events were causally determined by the operation of shuffling and it is only subjective ignorance that makes this appear random. Fair enough, but then substitute the shoes with two quantum entangled particles and you will convince yourself that you have ``determinism'' (in the sense of perfect correlations) without causality (see \cite{bell}).}

Similarly, take the digits of a (computable) number, say $\pi$. If you know with certainty that the number you are dealing with is really $\pi$, for instance because is the ratio between a circle's circumference and its diameter, then you can assert with probability one, i.e. deterministically, that its nineteenth decimal digit is a 4. Again, nobody would claim that you caused this digit to be 4 by measuring the circumference and diameter of a circle.

\section{Concluding remarks}
In this essay, we have revised arguments to support the view that classical physics could be interpreted indeterministically, and basic operational principles and information-theoretic arguments hint at this direction. At the same time, quantum mechanics has given us reason (in particular by means of the violation of \emph{Bell's inequalities} \cite{bell}) to believe that the universe we live in is not deterministic. If this is the case, Popper's words remind us of what is the reward for indeterminism: ``The future is open. It is not predetermined and thus cannot be predicted --except by accident. The possibilities that lie in the future are infinite'' \cite{poppermyth}. In fact, not  only is the future unpredictable in an indeterministic universe, but also the truth values of future (scientific) statements are genuinely \emph{undecidable}, as Gisin's simple example points out \cite{gisin2}:
\begin{displayquote}
Think of a proposition about the future, for example, ``It
will be raining in exactly one year time from
now at Piccadilly Circus''. If one believes in
determinism, then this proposition is either
true or false [...].
But if one believes that the future is open,
then it is not predetermined that it will rain,
hence the proposition is not true, and it is
not predetermined that it will not rain, thus
the proposition is also not false. 
\end{displayquote} 
However, our stance on an open future cannot remain but a belief, because compelling arguments (e.g., \cite{suppes, wendl}) show that every physical theory, including classical and quantum mechanics, can be interpreted either deterministically or indeterministically and no experiment will ultimately discriminate between these two opposite worldviews. We can only have the certainty that the future of the battle between determinism and indeterminism is open, too. 

\subsection*{Acknowledgments}
I would like to show my gratitude to Nicolas Gisin and Borivoje Daki\'c for the many interesting discussions and their comments that helped to improve this essay. I am indebted also to the many, interesting discussions held on the forum of FQXi Community during the 2020 Essay Contest.

\newpage

\begin{small}

\end{small}

\end{document}